\documentclass[journal]{IEEEtran}
\usepackage{amsmath}
\usepackage{graphicx}
\usepackage{url}

\newcommand{\sysmat}{A}
\newcommand{\im}{u}
\newcommand{\sino}{b}
\newcommand{\norm}[1]{\|#1\|}
\newcommand{\normone}[1]{\norm{#1}_1}
\newcommand{\normtwo}[1]{\norm{#1}_2}
\newcommand{\normtv}[1]{\norm{#1}_{\mathrm{TV}}}

\newcommand{\regpar}{\lambda}
\newcommand{\diffop}{D}
\DeclareMathOperator*{\argmin}{argmin}

\newcommand{\imtvsol}{\im_\mathrm{TV}}
\newcommand{\objfun}{f}

\newcommand{\rayindex}{i}
\newcommand{\pixelindex}{j}

\newcommand{\imdim}{2048}
\newcommand{\funname}{l1tv}
\newcommand{\alphaone}{2e-02}
\newcommand{\alphatwo}{2e-03}
\newcommand{\alphathree}{2e-04}

\newcommand{\half}{0.49}
\newcommand{\third}{0.32}
\newcommand{\mfac}{1.0}

\newcommand{\ee}[2]{#1 \cdot 10^{#2}}

\begin{document}
\title{Ensuring convergence in total-variation-based reconstruction for accurate microcalcification imaging in breast X-ray CT}

\author{Jakob~H.~J{\o}rgensen,~\IEEEmembership{Student Member,~IEEE,}
        Emil~Y.~Sidky,~\IEEEmembership{Member,~IEEE,}
        and~Xiaochuan~Pan,~\IEEEmembership{Fellow,~IEEE}
\thanks{
This work is part of the project CSI: Computational Science
in Imaging, supported by grant 274-07-0065 from the Danish
Research Council for Technology and Production Sciences. This work was supported in part by NIH R01 grants CA120540, EB000225.
The contents of this article are
solely the responsibility of the authors and do not necessarily
represent the official views of the National Institutes of Health.}
\thanks{J. H. J{\o}rgensen is with the Department of Informatics and Mathematical Modelling, Technical University of Denmark, Richard Petersens Plads, Byg\-ning 321, 2800 Kongens Lyngby, Denmark (e-mail: jakj@imm.dtu.dk).}%
\thanks{E. Y. Sidky and X. Pan are with the Department of Radiology, University of Chicago, 5841 S. Maryland Ave., Chicago IL, 60637, USA (e-mail: \{sidky,xpan\}@uchicago.edu).}}%

\maketitle
\pagestyle{empty}
\thispagestyle{empty}

\begin{abstract}
Breast X-ray CT imaging is being considered in screening as an extension to mammography. As a large fraction of the population will be exposed to radiation, low-dose imaging is essential. Iterative image reconstruction based on solving an optimization problem, such as Total-Variation minimization, shows potential for reconstruction from sparse-view data. For iterative methods it is important to ensure convergence to an accurate solution, since important image features, such as presence of microcalcifications indicating breast cancer, may not be visible in a non-converged reconstruction, and this can have clinical significance. To prevent excessively long computational times, which is a practical concern for the large image arrays in CT, it is desirable to keep the number of iterations low, while still ensuring a sufficiently accurate reconstruction for the specific imaging task. This motivates the study of accurate convergence criteria for iterative image reconstruction. In  simulation studies with a realistic breast phantom with microcalcifications we compare different convergence criteria for reliable reconstruction. Our results show that it can be challenging to ensure a sufficiently accurate microcalcification reconstruction, when using standard convergence criteria. In particular, the gray level of the small microcalcifications may not have converged long after the background tissue is reconstructed uniformly. We propose the use of the individual objective function gradient components to better monitor possible regions of non-converged variables. For microcalcifications we find empirically a large correlation between nonzero gradient components and non-converged variables, which occur precisely within the microcalcifications. This supports our claim that gradient components can be used to ensure convergence to a sufficiently accurate reconstruction.
\end{abstract}


\section{Introduction}
%
%
%
\IEEEPARstart{D}{ose} reduction has gained considerable interest in diagnostic computed tomography (CT) in recent years \cite{McCollough:09}.
The potential to employ CT for screening, where
a large population fraction will be exposed to radiation dose and
the majority of subjects will be asymptomatic, also motivates the interest in low intensity X-ray CT.
Breast CT poses a particularly challenging problem as the total exposure is restricted to the equivalence of two digital mammograms. Such a low X-ray dose can be achieved either by drastically reducing the intensity compared to a diagnostic-quality CT scan, or by reconstruction from sparse-view data.
%
%
%
%
%
%
%
%
%

Total-Variation (TV)-regularized image reconstruction exploits approximate sparsity of the spatial gradient of cross sections of the human body to compensate for reduction in data. TV-reconstructions have been shown to compare favorably with standard Filtered Back Projection from sparse-view data \cite{Sidky:06, Sidky:08}.
We are investigating the optimal trade-off between low intensity views and sparse-view data for breast CT by means of TV-reconstruction \cite{Joergensen:2011}.
%

The TV-reconstruction is obtained by solving a nonlinear optimization problem.
A practical concern is that the extremely large systems in CT, where image arrays of $10^9$ voxels are standard, are challenging to solve accurately in acceptable time. Complicating this issue is the fact that clinically relevant features are often very small---occupying only a few voxels. As result both global and pointwise convergence of an iterative reconstruction algorithm may have clinical impact. We demonstrate this issue in the present preliminary investigation, where we 
examine a realistic simulation of CT for breast cancer screening, and compare strategies for ensuring convergence to a sufficiently accurate TV-reconstruction.

\section{Image reconstruction by TV-minimization}
We consider TV-regularized image reconstruction in order to exploit gradient sparsity to compensate for the few-view projection data. The present study works with the discrete-to-discrete imaging model, $\sysmat \im = \sino$,  see \cite{Barrett:FIS}.
%
For reconstruction we consider the minimization problem
\begin{align}
 \imtvsol &= \argmin_\im \objfun(\im), \label{eq:minprob}
\end{align}
where
\begin{align}
 \objfun(\im) = \normone{\sysmat\im-\sino} + \regpar \normtv{\im} \label{eq:tvrecmodel}
\end{align}
and
\begin{align}
 \normtv{\im} = \sum_\pixelindex \normtwo{\diffop_\pixelindex \im}
\end{align}
and $\diffop_\pixelindex$ is a forward difference approximation to the image gradient at pixel $\pixelindex$.

Instead of the more commonly used $\ell_2$ norm for measuring data fidelity we use the $\ell_1$ norm. TV-regularized $\ell_2$ norm minimization is known to be contrast-reducing, in particular for objects of small scale \cite{Strong:03}, such as microcalcifications. $\ell_1$ minimization does not remove this problem, but tends to reduce it \cite{Chan:05}. 

Both terms in \eqref{eq:tvrecmodel} are non-differentiable, and in order to apply standard gradient-based optimization algorithms we apply the standard smoothing trick of the replacements:
\begin{align}
 \sum_\rayindex \sqrt{\normtwo{\diffop_\pixelindex \im }^2 + \epsilon} \qquad &\text{replaces} \qquad \normtv{\im}.\\
 \sum_\rayindex \sqrt{|(\sysmat\im)_\rayindex-\sino_\rayindex|^2 + \epsilon} \qquad &\text{replaces} \qquad \normone{\sysmat\im-\sino}.
\end{align}
In our simulations we use $\epsilon = 10^{-4}$, which we found sufficiently small to prevent any change in visual appearance of the reconstructed image compared to using $\epsilon = 0$.

\begin{figure}[t]
\begin{center}
\begin{minipage}{\half\linewidth}
\pdfximage width \linewidth {orig_n_\imdim.png}
\pdfrefximage\pdflastximage 
\end{minipage}
\begin{minipage}{\half\linewidth}
\includegraphics[width=\linewidth]{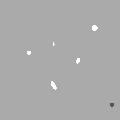}  
\end{minipage}
\end{center}
\caption{Left: Original full breast phantom, $\imdim^2$ pixels. Right: $120^2$ pixel ROI around simulated microcalcifications. Gray level window: $[0.9, 1.2]$.\label{fig:origphantom}}
\end{figure}
\begin{figure}[htb]
 \begin{minipage}{\half\linewidth}
\pdfximage width \textwidth {rec_\funname_n_\imdim_alpha_\alphaone.png}
\pdfrefximage\pdflastximage
\pdfximage width \textwidth {rec_\funname_n_\imdim_alpha_\alphatwo.png}
\pdfrefximage\pdflastximage
\pdfximage width \textwidth {rec_\funname_n_\imdim_alpha_\alphathree.png}
\pdfrefximage\pdflastximage
\end{minipage}
\begin{minipage}{\half\linewidth}
\includegraphics[width=\textwidth]{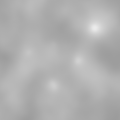} 
 \includegraphics[width=\textwidth]{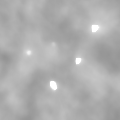} 
 \includegraphics[width=\textwidth]{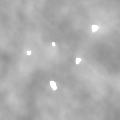} 
\end{minipage}
\caption{Reconstructions of full image and ROIs. Top: $\regpar=2\cdot 10^{-2}$, middle: $\regpar=2\cdot 10^{-3}$, bottom: $\regpar=2\cdot 10^{-4}$. Gray level window: $[0.9, 1.2]$. \label{fig:recs}}
\end{figure}

An important question is how well a TV reconstrution is capable of reproducing the salient image features, such as microcalcifications in the present case. Numerous studies demonstrate that of TV-reconstruction can produce clinically useful reconstructions, see e.g. \cite{Sidky:06,Sidky:08}. 


Our main question of interest in the present work arises when using an iterative algorithm to solve the TV minimization problem: When can we reliably stop iterating and accept the computed solution as a good approximation of the true minimizer to \eqref{eq:tvrecmodel}? In other words, what is a good termination criterion?

Note that, in general, $\imtvsol$ is biased compared to the original underlying image, and the size of this bias is parameter dependent, in particular the bias depends on $\regpar$. It is not our goal to select a well-suited $\regpar$ here; we only consider the question of, given a choice of $\regpar$, how do the iterates approach the solution, i.e., the minimizer of \eqref{eq:tvrecmodel}? We will consider two different choices for termination criterion for the iterative algorihtm used for solving \eqref{eq:minprob}:
\begin{enumerate}
 \item $\normtwo{\nabla \objfun(\im)} < \tau$
 \item $1 + \cos \alpha < \tau$,
\end{enumerate}
where $\alpha$ is the angle between the gradients of each of the two terms in \eqref{eq:tvrecmodel}, see the original reference \cite{Sidky:08} for details, and $\tau$ is a user-specified tolerance, where a smaller $\tau$ leads to a more accurate solution. Both criteria correspond to theoretical optimality conditions \cite{Nocedal:2006} in the limit of $\tau = 0$.

For solving \eqref{eq:minprob} we use a convergent, gradient-based optimization algorithm, which is optimal in a certain sense, see \cite{Jensen}. The algorithm was developed for TV-regularized $\ell_2$ data fidelity, but is applicable to any smooth objective function, and we have found that it works well for solving (the smoothed version of) the problem in \eqref{eq:minprob}.

\section{Breast CT model}
Breast CT imaging is being considered as a potential addition to mammography in screening for breast cancer. One particular indicator of breast cancer is formation of \emph{microcalcifications}---very small, highly attenuating calcium deposits. For screening, low-dose imaging is pertinent to minimize accumulated X-ray dose, while accurate and reliable microcalcification shape and attenuation reconstruction may be important for detecting malignancy.

\begin{figure*}[t]
\begin{center}
\begin{minipage}{\third\textwidth}
 \includegraphics[width=\mfac\linewidth]{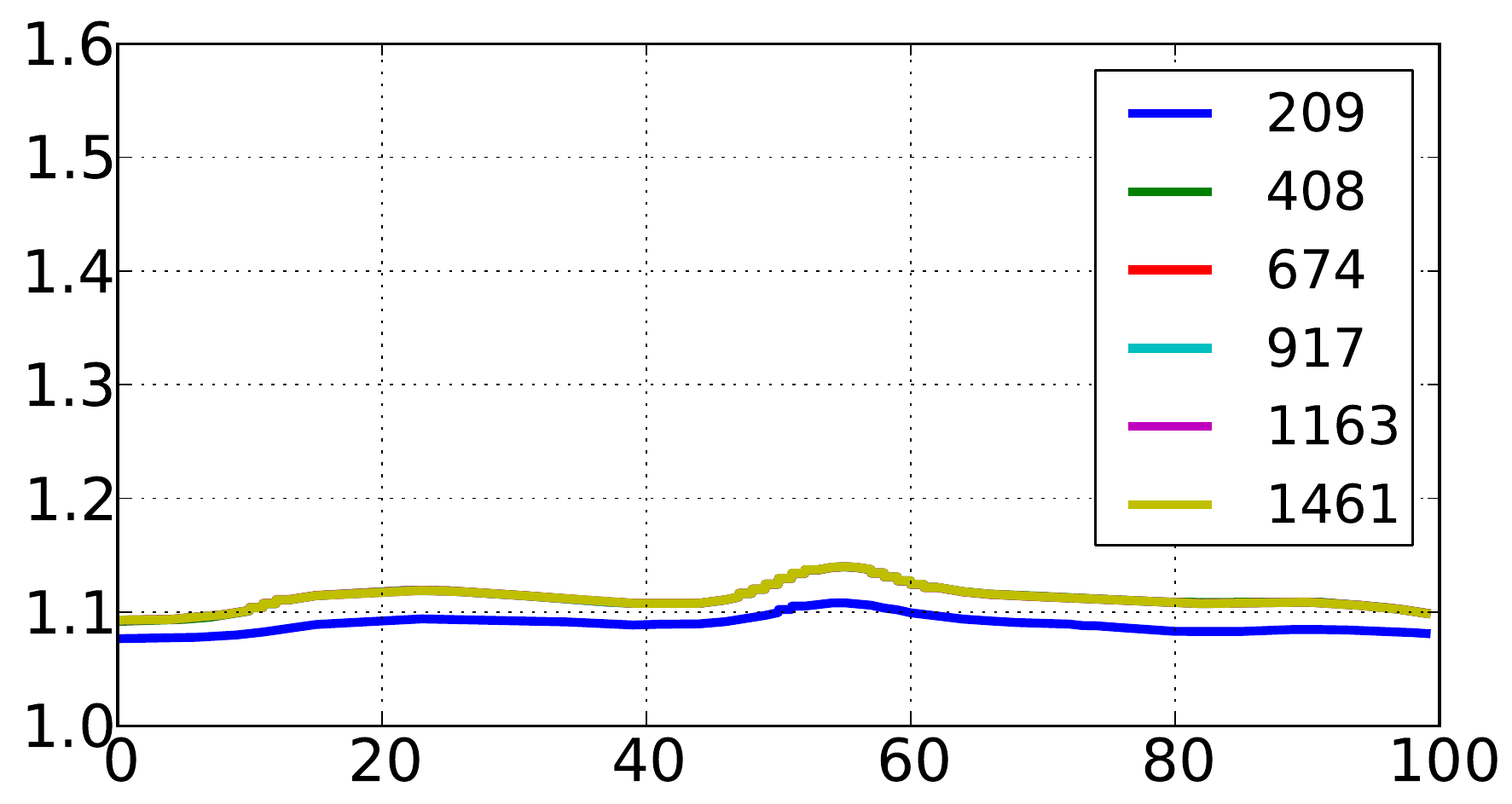} 
 \includegraphics[width=\mfac\textwidth]{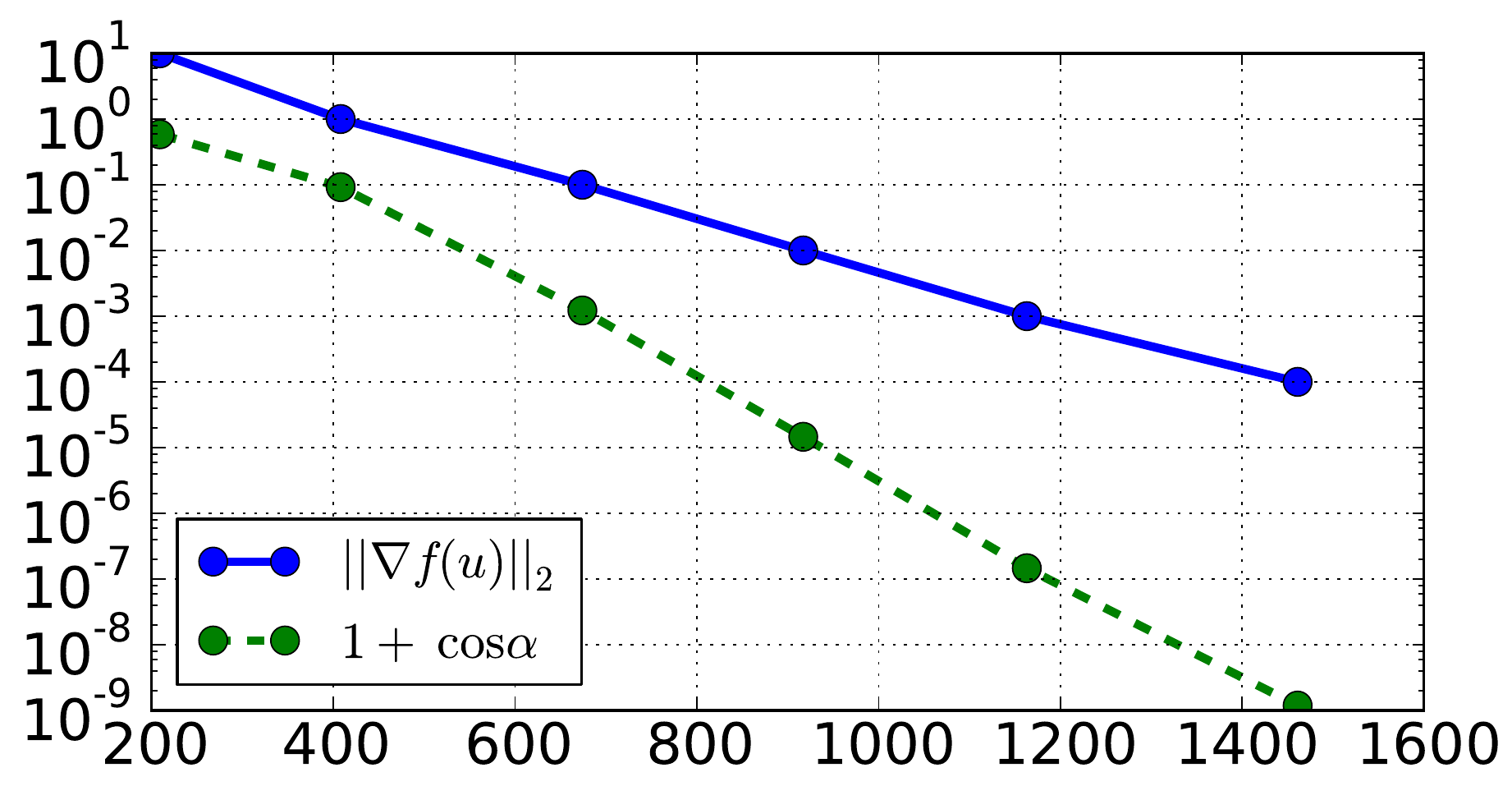} 
\includegraphics[width=\mfac\textwidth]{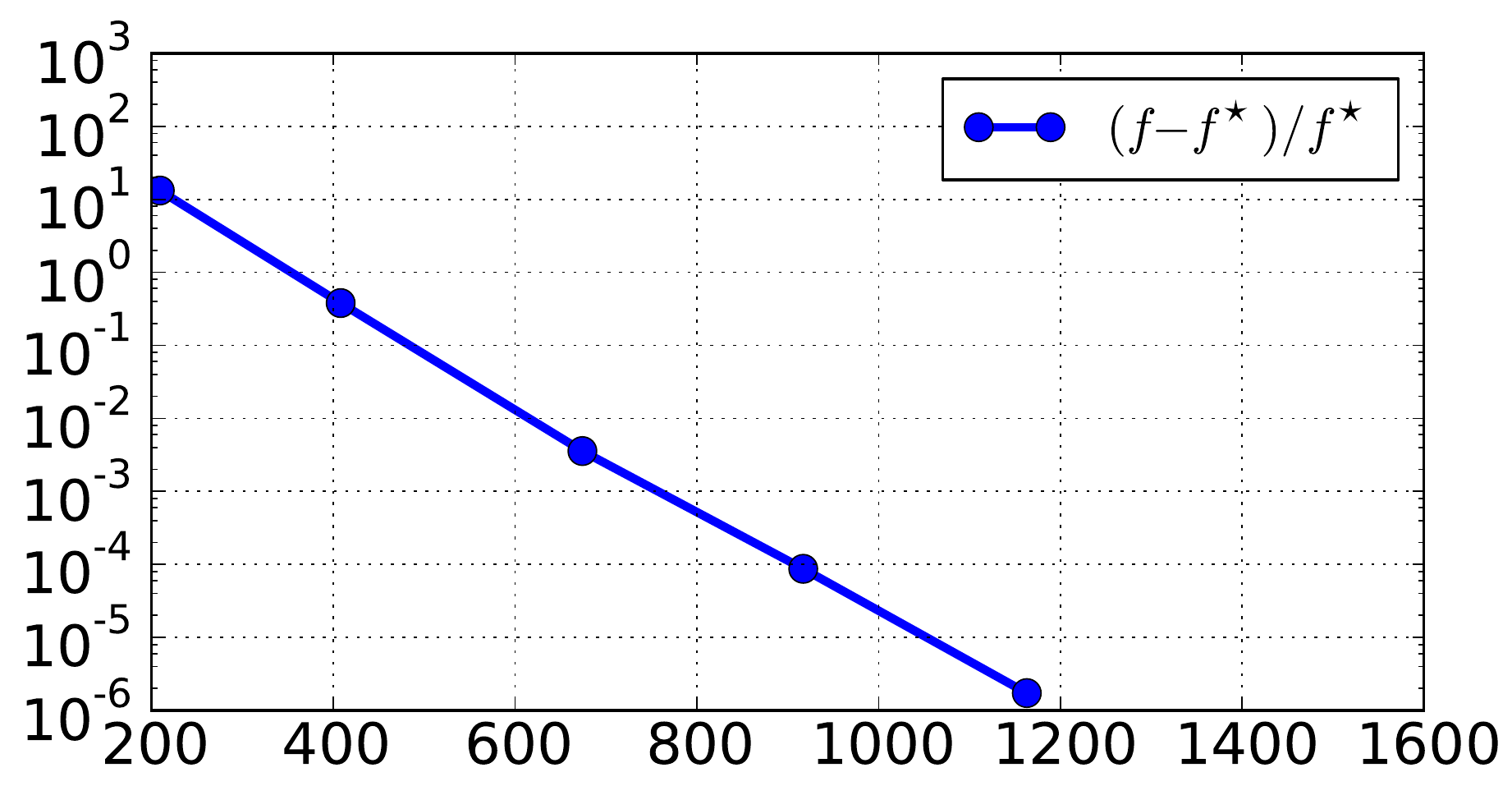} 
\end{minipage}
\begin{minipage}{\third\textwidth}
 \includegraphics[width=\mfac\linewidth]{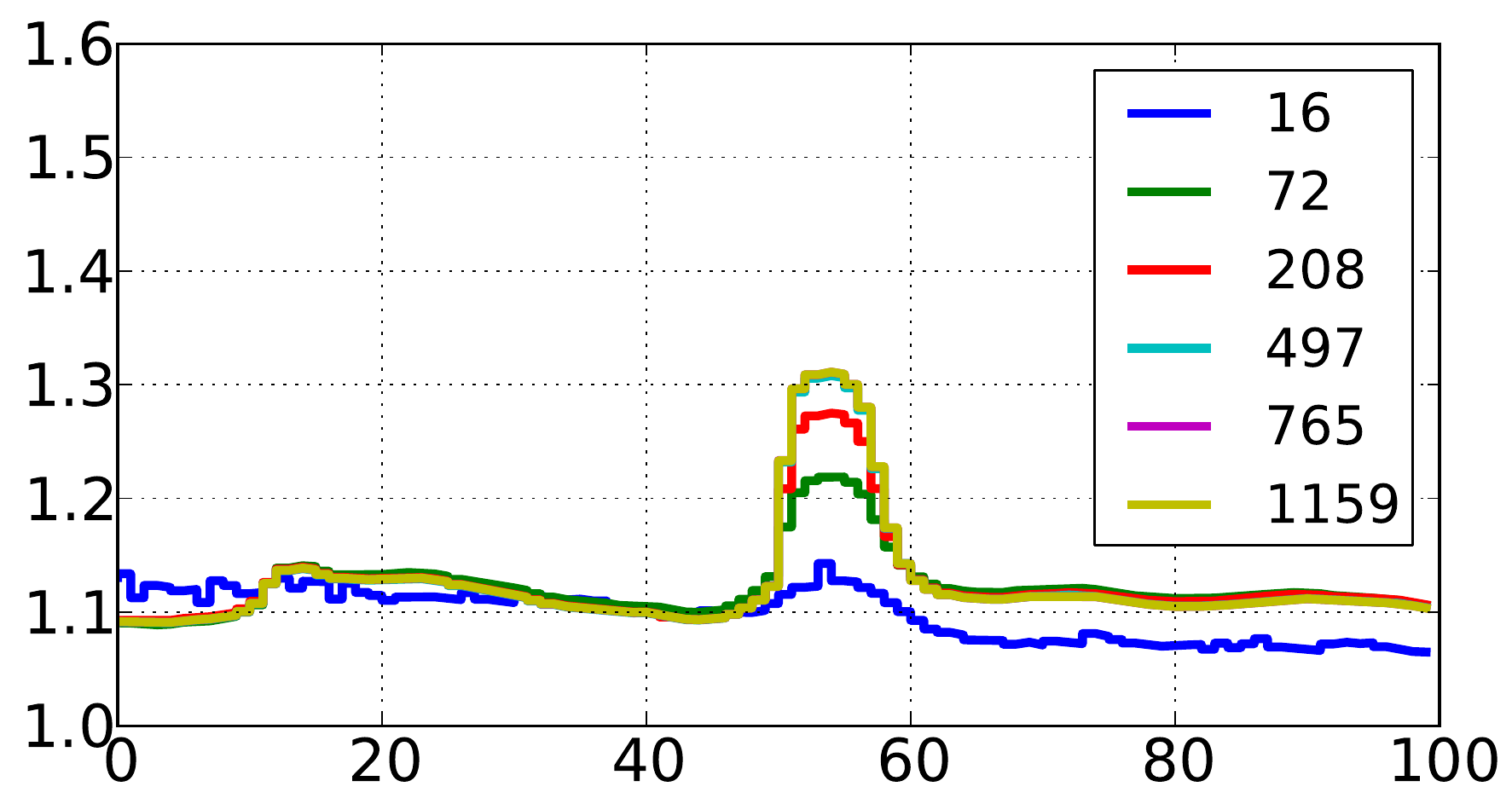} 
 \includegraphics[width=\mfac\textwidth]{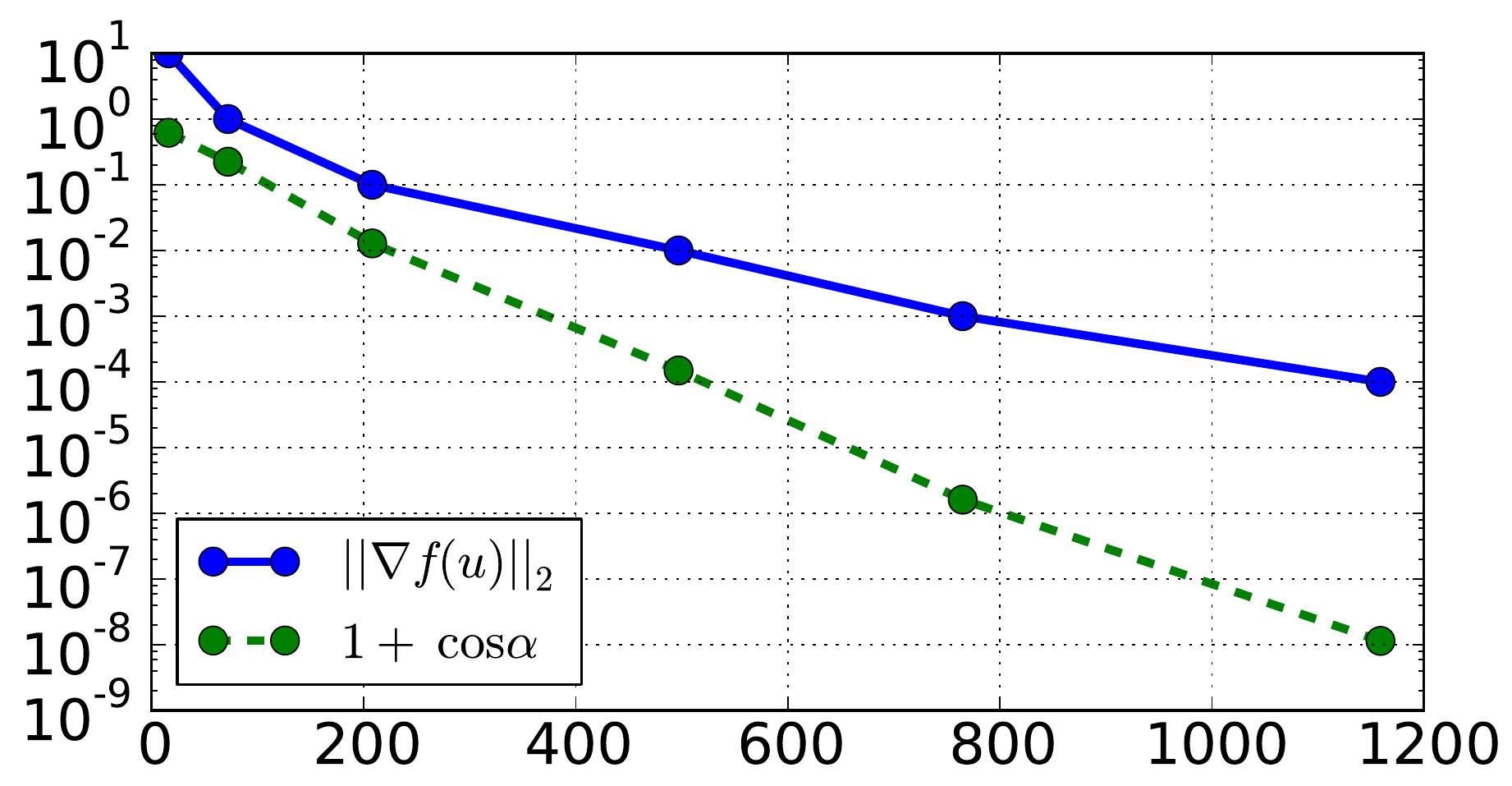} 
\includegraphics[width=\mfac\textwidth]{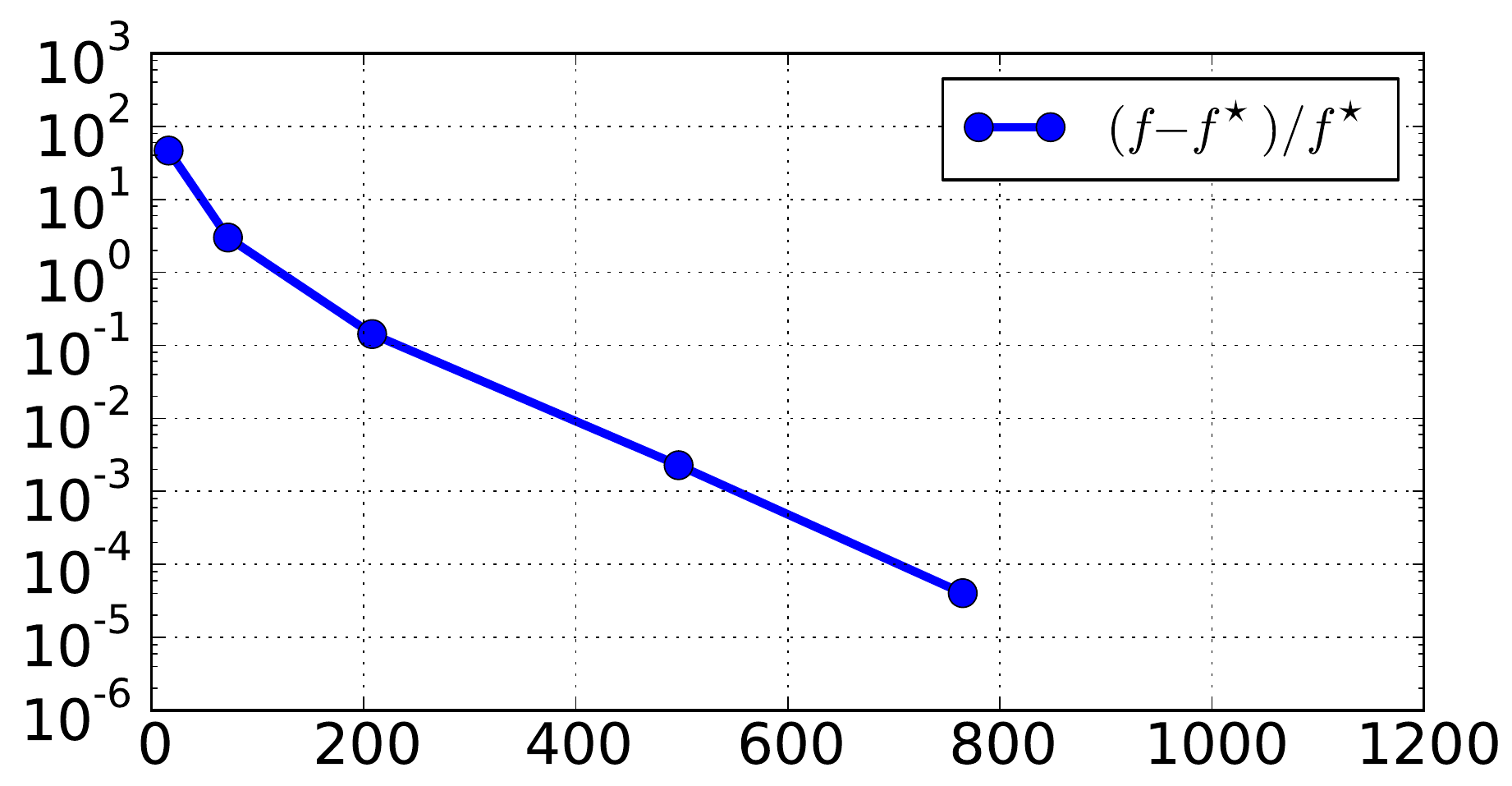} 
\end{minipage}
\begin{minipage}{\third\textwidth}
 \includegraphics[width=\mfac\linewidth]{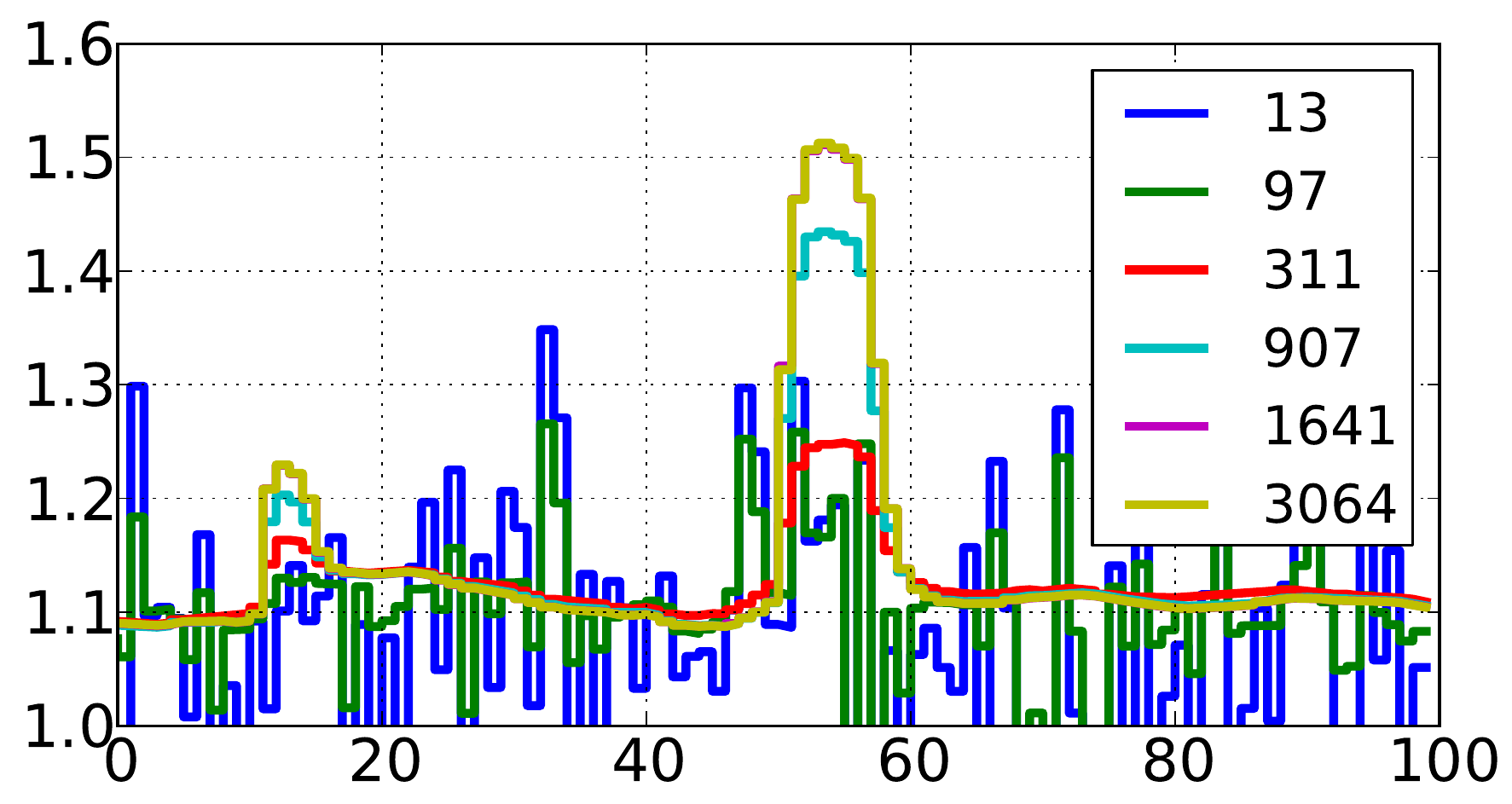} 
 \includegraphics[width=\mfac\textwidth]{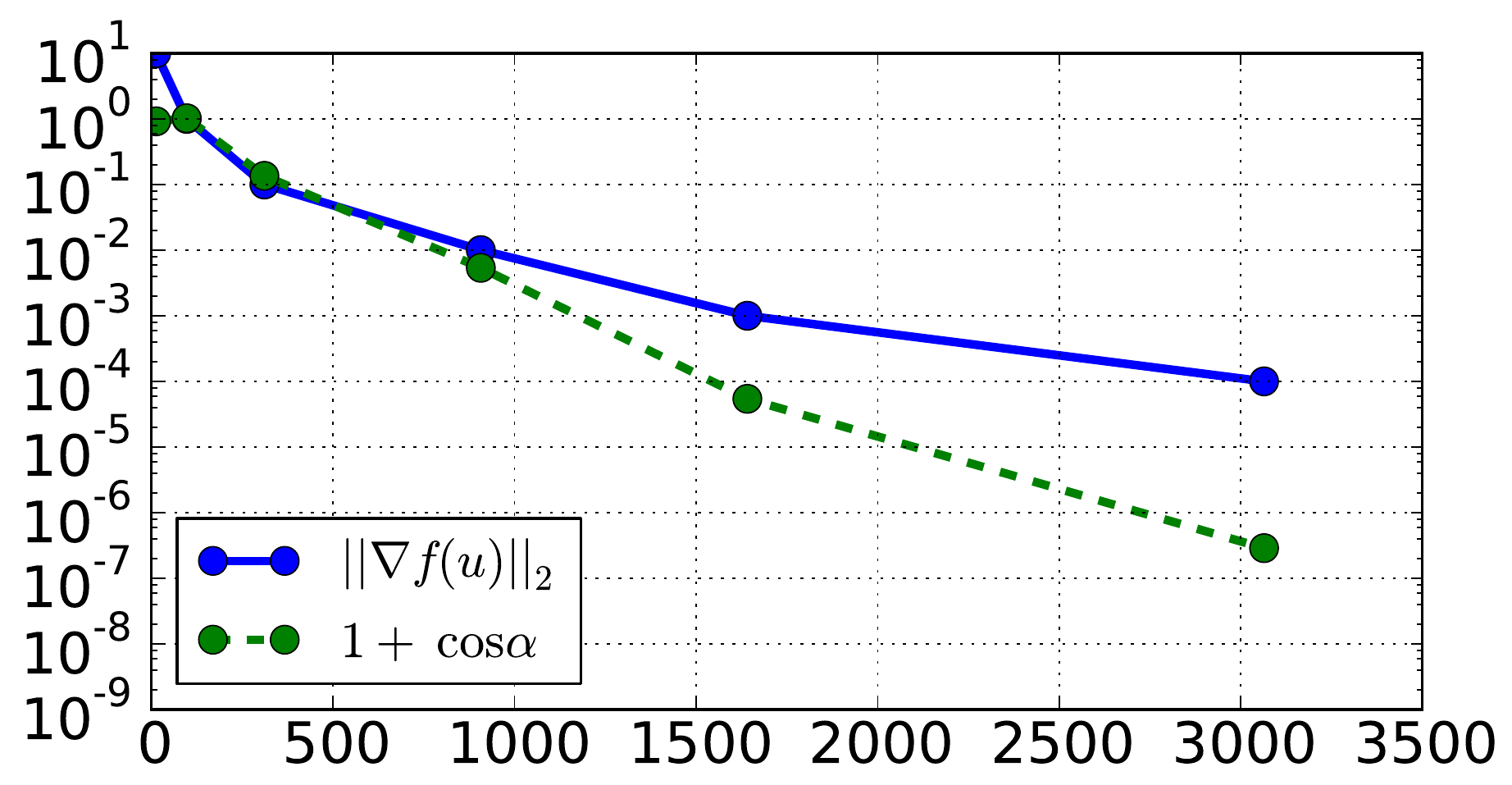} 
\includegraphics[width=\mfac\textwidth]{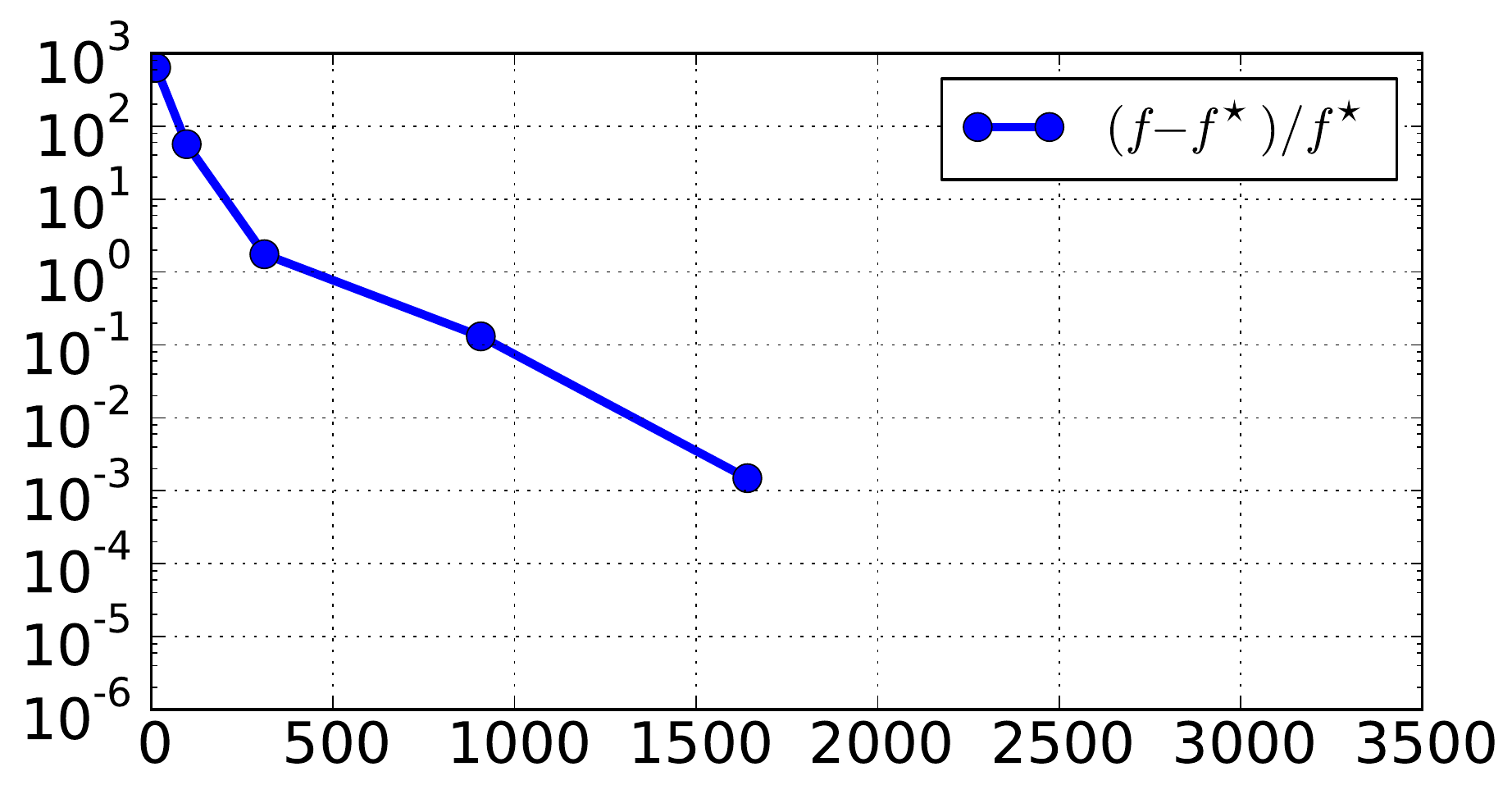} 
\end{minipage}
\caption{Top row: Vertical profile through microcalcifications for iteration number corresponding to terminating iterations at $\tau = 10^1, 10^0, \dots, 10^{-4}$ using criterion 1. Middle row: Values of the two convergence criteria vs. number of iterations. Bottom row: Convergence in objective function relative to the reference solution vs. number of iterations. Left column: $\regpar = 2\cdot 10^{-2}$, center column: $\regpar = 2\cdot 10^{-3}$, right column: $\regpar = 2\cdot 10^{-4}$. \label{fig:profile_conv}}
\end{center}
\end{figure*}

In the present work we use the breast phantom from \cite{reiser2010task} discretized on a $\imdim^2$ pixel grid, as shown in Fig. \ref{fig:origphantom} along with a $120^2$ pixel region of interest (ROI) around a simulated cluster of microcalcifications, also discretized. Gray values in units of water attenuation are given in Table \ref{tab:graylevels}. Note the fairly complex phantom structure, which makes the phantom semi-realistic, and at the same time poses a challenge for TV-based reconstruction, which tends to favor piecewise constant, ``cartoon-like'' images.

\begin{table}[htb]
\centering
 \begin{tabular}{|l|c|}
\hline
 Tissue & Value \\
\hline
Fat & $1.00$ \\
Fibro-glandular tissue & $1.10$ \\
Skin & $1.15$ \\
Microcalcifications & $1.80 - 2.10$\\
\hline
\end{tabular}
\caption{Gray values for breast phantom, in units of water attenuation.\label{tab:graylevels}}
\end{table}


\section{Numerical results} \label{sec:numerical}
Different choices for the regularization parameter $\regpar$ lead to very different solutions, and the question of how to choose a well-suited $\regpar$ is important. However, our goal here is merely to demonstrate that very different convergence is observed for different choices of $\regpar$; not to propose a certain $\regpar$ over others. For that purpose we make three choices: $\alpha = 2\cdot 10^{-2}$, $\alpha = 2\cdot 10^{-3}$, and $\alpha = 2\cdot 10^{-4}$. We generate noise-free $64$-view, $1024$-detector-bin fan-beam data by forward projection (using a line intersection-based ray-driven projector) of the original discrete $2048^2$ pixelized phantom with microcalcifications. We solve \eqref{eq:minprob} with termination criterion 1 for $\tau = 10^{-4}$ to obtain accurate solutions. The obtained reconstructions are shown in Fig. \ref{fig:recs}.

As expected, with increasing $\regpar$ the reconstructed images becomes smoother, and the microcalcifications gradually become invisible. Only at $\regpar = 2\cdot 10^{-4}$ is the smallest microcalcification visible, so it is clear that we need to use a $\regpar$ smaller than or equal to $2\cdot 10^{-4}$.

We rerun the three reconstructions and store this time iterates along the way, at thresholds $\tau = 10^1, 10^0, \dots, 10^{-4}$ for termination criterion 1. We use the most accurate iterate, at each $\regpar$ as a \emph{reference solution} for comparing the convergence of the earlier iterates. We denote the reference solution by $\im^\star$ and its value for the objective function by $\objfun^\star$.

In the top row in Fig. \ref{fig:profile_conv} we show reconstruction profiles through two of the microcalcifications (the two that are on the same vertical line) for each of the stored iterates, including the reference solutions. For the largest $\regpar$ we see that the iterates converge to the reference solution very quickly: after $438$ iterations i.e. at $\tau = 10^0$ the solution is indistinguishable from the reference solution. For the middle $\regpar$ we see a different behavior to which we will refer as \emph{non-uniform convergence}: For the most part, the iterates converge to the reference solution rapidly, but precisely within the larger of the two microcalcifications, the iterates converge very slowly. For the smallest $\regpar$ the non-uniform convergence is even more pronounced and only at the reconstruction stored before the reference solution we see no further improvement of the iterates. It seems natural that for even smaller values of $\regpar$ we would see even more severe non-uniform convergence.

\begin{figure*}[t]
\begin{minipage}{0.03\textwidth}
\begin{picture}(0,0)(0,0)
 \put(0,95){$\tau$}
\put(0,50){$10^{-1}$}
\put(0,-20){$10^{-2}$}
\put(0,-90){$10^{-3}$}
\end{picture}
\end{minipage}
\begin{minipage}{0.45\textwidth}
 \centering Difference images to pseudo-solution\\
\begin{minipage}{\third\textwidth}
$$\lambda = 2\cdot 10^{-2}$$
 \includegraphics[width=\textwidth]{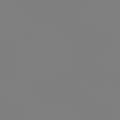} 
\includegraphics[width=\textwidth]{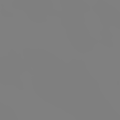} 
\includegraphics[width=\textwidth]{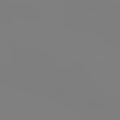} 
\end{minipage}
\begin{minipage}{\third\textwidth}
$$\lambda = 2\cdot 10^{-3}$$
 \includegraphics[width=\textwidth]{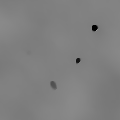} 
\includegraphics[width=\textwidth]{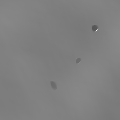} 
\includegraphics[width=\textwidth]{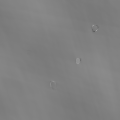} 
\end{minipage}
\begin{minipage}{\third\textwidth}
$$\lambda = 2\cdot 10^{-4}$$
 \includegraphics[width=\textwidth]{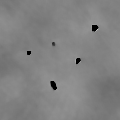} 
\includegraphics[width=\textwidth]{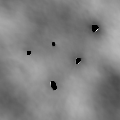} 
\includegraphics[width=\textwidth]{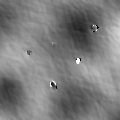} 
\end{minipage}
\end{minipage}
\begin{minipage}{0.01\textwidth}
\;
\end{minipage}
\begin{minipage}{0.45\textwidth}
  \centering Gradient components\\
 \begin{minipage}{\third\textwidth}
$$\lambda = 2\cdot 10^{-2}$$
 \includegraphics[width=\textwidth]{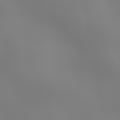} 
\includegraphics[width=\textwidth]{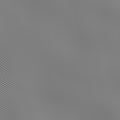} 
\includegraphics[width=\textwidth]{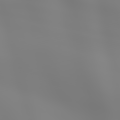} 
\end{minipage}
\begin{minipage}{\third\textwidth}
$$\lambda = 2\cdot 10^{-3}$$
 \includegraphics[width=\textwidth]{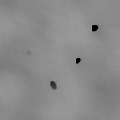} 
\includegraphics[width=\textwidth]{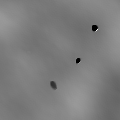} 
\includegraphics[width=\textwidth]{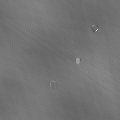} 
\end{minipage}
\begin{minipage}{\third\textwidth}
$$\lambda = 2\cdot 10^{-4}$$
 \includegraphics[width=\textwidth]{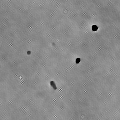} 
\includegraphics[width=\textwidth]{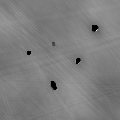} 
\includegraphics[width=\textwidth]{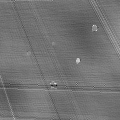} 
\end{minipage}
\end{minipage}
\caption{Left: Difference images $\im - \im^\star$. Gray level windows: Top: $[\ee{-1}{-1}, \ee{1}{-1}]$. Middle: $[\ee{-2}{-2}, \ee{2}{-2}]$. $[\ee{-2}{-3}, \ee{2}{-3}]$. Right: Gradient components $\nabla \objfun(\im)$. Gray level windows: Top: $[\ee{-1}{-4}, \ee{1}{-4}]$. Middle: $[\ee{-1}{-5}, \ee{1}{-5}]$. $[\ee{-1}{-6}, \ee{1}{-6}]$. \label{fig:diffims_gradcomps}}
\end{figure*}

Our concern about non-uniform convergence arises from two facts: First, detecting non-uniform convergence can be very challenging as we will demonstrate. Second, if we are not aware of non-uniform convergence, we risk accepting a solution which is not yet converged everywhere. Such a reconstruction has much lower contrast than the true TV-solution, which will make it difficult to spot the microcalcifications. This can lead us to the, incorrect, conclusion that the TV-solution is not capable of reproducing microcalcifications faithfully, but in fact the lack of contrast in the reconstruction was a result of accepting a too early iterate returned by the iterative solver and not because of the TV-minimization problem itself.

To illustrate that detecting non-uniform convergence is challenging, we compare the use of termination criteria 1 and 2. In the middle row of plots in Fig. \ref{fig:profile_conv} we have plotted the values of the two criteria vs. the number of iterations used for the considered iterates . Furthermore, in the bottom row of plots we show the convergence in terms of objective function value relative to the reference solution vs. the number of iterations. The difference in function value relative to the reference solution acts as an indicator of the accuracy of the iterate. Now imagine that we use $\tau = 10^{-3}$ as our convergence criterion. With criterion 1 and the largest $\regpar$ we find an $(\objfun - \objfun^\star)/\objfun^\star$ of approx. $10^{-6}$, which we consider to be very accurate. However, for the two smaller values $\regpar$ we find  $(\objfun - \objfun^\star)/\objfun^\star$ of approx. $10^{-4}$ and $10^{-3}$, indicating much less accurate reconstruction. A similar trend can be observed for termination criterion 2. The different accuracies obtained confirm our observations from the profile plots in the top row. We note that we are able to detect non-uniform by comparing the final function value differences in the plots in the bottom row. However, in practice, we do not have access to the true solution or a reference solution, as we would like keep the number of iterations low. We do have access to the values of the termination criterion functions, but as can be seen by inspecting the middle row of plots, we cannot trust that a using a fixed $\tau$ will provide a uniformly converged reconstruction.

\section{Gradient components}

As a first step towards a more reliable convergence criterion we wish to point out a connection that can possibly exploited. The two considered convergence criteria both involve the gradient of the objective function $\objfun$. However, as we saw, they do not clearly show that a few pixels have not yet reached convergence. We believe this is due to computing a single number from the full gradient for comparing with a $\tau$, thereby ``averaging out'' the differences between the individual components of the gradient. Many small gradient components will tend to hide the presence of a few larger ones. We propose instead to monitor the full objective function gradient $\nabla \objfun(\im)$ during the iterations.

In the right half of Fig. \ref{fig:diffims_gradcomps} we display as an image the ROI gradient components of the objective function, for the iterates obtained with $\tau = 10^{-1}$, $10^{-2}$, $10^{-3}$ for each of the three choices of $\regpar$. In the left half of Fig. \ref{fig:diffims_gradcomps} we show the corresponding ROI difference images $\im - \im^\star$ between the iterates and the reference solution image. Note that with decreasing $\tau$ we are making the gray level windows narrower to emphasize small components. 

For the largest $\regpar$ both difference images and gradient components appear to be zero (in the chosen gray level window) for all three choices of $\tau$. This agress well with our observatoin that the iterates converged rapidly to the reference solution, so we should precisely expect very small gradient components everywhere here. 

For the smaller choices of $\regpar$ we observe a highly non-uniform nonzero gradient component pattern for both the difference images and the gradient components, with large (negative) components exactly at the microcalcifications. The gradient components are negative, which agress with the variables still growing as seen in the profiles in the top row of Fig. \ref{fig:profile_conv}. For the more accurate reconstructions the microcalcification pixel values in the difference image and gradient components remain distinct while their magnitude approach zero. 

There is a clear correlation between the difference images and the gradient components, indicating a close connection. This suggests the possibility for ensuring local convergence in the microcalcifications by means of monitoring the gradient components. 

At the most accurate solution, while the microcalcification pixels are still visible in difference images and gradient components, the intensity is of the level of the background. This leads us to the conclusion that at this point the iterate has converged, and we can relibly accept it as an accurate solution.

\section{Discussion}
Note that we are only able to monitor the convergence using the difference images, since we computed the reference solution. In practice, we wish to monitor convergence at any given iteration without a much more accurate reference solution. The gradient components are readily available during the iterations, and as our simulation shows, they can be used to monitor non-converged pixels. 

We are investigating strategies other than visual inspection of the gradient components for a quantitative convergence criterion. For instance by forcing $\max_\pixelindex \vert (\nabla \objfun(\im))_\pixelindex \vert$ below a appropriately chosen threshold $\epsilon$, all gradient components will be smaller than $\epsilon$, thereby ensuring global convergence. 
When applying a single number based convergence criterion such criteria 1 and 2, the fact that the majority of the variables are at optimum can conceal by averaging out the contributions from the few variables that are not. The rationale in forcing all gradient components below $\epsilon$ is that small areas of non-convergent varibles will prevent termination of the algorithm.
A different approach would be to exploit the spatial structure in the nonzero gradient components, e.g. by not terminating iterations until no spatial correlation is present.

\section{Conclusion}
We have conducted a preliminary comparative investigation of convergence criteria for ensuring accurate reconstruction of microcalcifications in breast CT. We have demonstrated that the nonzero gradient components can be used to monitor the regions of non-converged variables and thereby preventing termination of the optimization algorithm before global convergence is reached.

Accepting a reconstruction which is not globally converged may have clinical significance, for instance, as in the example given, by providing insufficient contrast for detecting the microcalcifications.

The use of the objective gradient in a convergence criterion is well-known, at least the use of the norm of the gradient. Explicit use of the individual gradient components for monitoring local convergence for small objects such as microcalcifications has not, to the best of our knowledge, been studied before. An interesting direction for future work is to apply the approach to other optimization based reconstruction techniques.

\bibliographystyle{IEEEtran}
\bibliography{mic2011_CR_references}

\end{document}